# Optimization of automatically generated multi-core code for the LTE RACH-PD algorithm

**Maxime Pelcat**
IETR/INSA, UMR CNRS 6164,
Rennes, France
mpelcat@insa-rennes.fr

**Slaheddine Aridhi**
Texas Instruments, CIV Division,
Villeneuve Loubet, France
saridhi@ti.com

**Jean-François Nezan**
IETR/INSA, UMR CNRS 6164,
Rennes, France
jnezan@insa-rennes.fr

*Abstract—* Embedded real-time applications in communication systems require high processing power. Manual scheduling developed for single-processor applications is not suited to multi-core architectures. The Algorithm Architecture Matching (AAM) methodology optimizes static application implementation on multi-core architectures.
The Random Access Channel Preamble Detection (RACH-PD) is an algorithm for non-synchronized access of Long Term Evolution (LTE) wireless networks. LTE aims to improve the spectral efficiency of the next generation cellular system. This paper describes a complete methodology for implementing the RACH-PD. AAM prototyping is applied to the RACH-PD which is modelled as a Synchronous DataFlow graph (SDF). An efficient implementation of the algorithm onto a multi-core DSP, the TI C6487, is then explained. Benchmarks for the solution are given.

## I. INTRODUCTION

The recent evolution of digital communication systems (voice, data and video) has been dramatic. Over the last two decades, low data-rate systems have been replaced or augmented by systems capable of data rates of several Mbit/s, supporting multimedia applications (such as DSL, cable modems, 802.11b/a/g/n wireless local area networks, 3G and WiMAX). The 3GPP Long Term Evolution (LTE) represents a recent part of this evolution, enabling data rates beyond hundreds of Mbit/s in potentially very wide cells.

As communication systems have evolved, the resulting increase in data rates has necessitated higher system algorithmic complexity. A more complex system requires greater flexibility in order to function with different protocols in diverse environments. Additionally, there is an increased need for the system to support multiple interfaces and multi-component devices. Consequently, this requires the optimization of device parameters over varying constraints, such as performance, area and power. Achieving this device optimization requires a good understanding of the application complexity and the choice of an appropriate architecture to support this application.

System on a Chip (SoC) with several cores such as multi-core DSPs is becoming the standard basic element used to build complex telecommunication systems. The task of distributing pieces of an algorithm over a multi-component architecture is not straightforward. When performed manually, the result is inevitably a sub-optimal solution. There is a need for new methodologies that allow the exploration of several solutions thus producing a more optimal result. For the current work, the methodology of Algorithm-Architecture Matching (AAM, previously called AAA [6]) is employed using the Parallel Real-time Embedded Executives Scheduling Method (PREESM) tool. The PREESM tool is an open framework which provides a flexible method for exploring architectures suited for deterministic applications. More than just a simulation tool, PREESM can generate code. Associated with well-optimized code, communication and synchronization, the automatic generation leads to an efficient algorithm implementation.

This article presents an overview of the LTE Random Access Channel (RACH) preamble detection algorithm and the PREESM tool. Subsequently, the preamble detection application is described using a Synchronous DataFlow graph (SDF). The virtual prototyping of this application over multi-processor architectures using PREESM tool features is then detailed. The target architecture is a multi-core DSP from Texas Instruments, the C6487. An implementation onto this DSP is performed with optimized inter-core communication and synchronizations using Direct Memory Access (DMA). Finally future work is discussed and conclusions are given.

## II. PREAMBLE DETECTION PROCESS

The RACH is a contention-based uplink channel used mainly for initial transmission requests from the User Equipment (UE) to the evolved base station (eNodeB) for connection to the network. The UE seeking connection with a base station sends its signature in a RACH preamble dedicated time and frequency window in accordance with a predefined preamble format. Signatures have special auto-correlation and inter-correlation properties that maximize the ability of the eNodeB to distinguish one UE from another. The RACH preamble procedure is implemented in the LTE eNodeB to detect and identify each user's signature and is dependent on the cell size and the system bandwidth. We assume that the eNodeB has the capacity to handle the processing of this RACH preamble detection every millisecond in a worst case scenario.

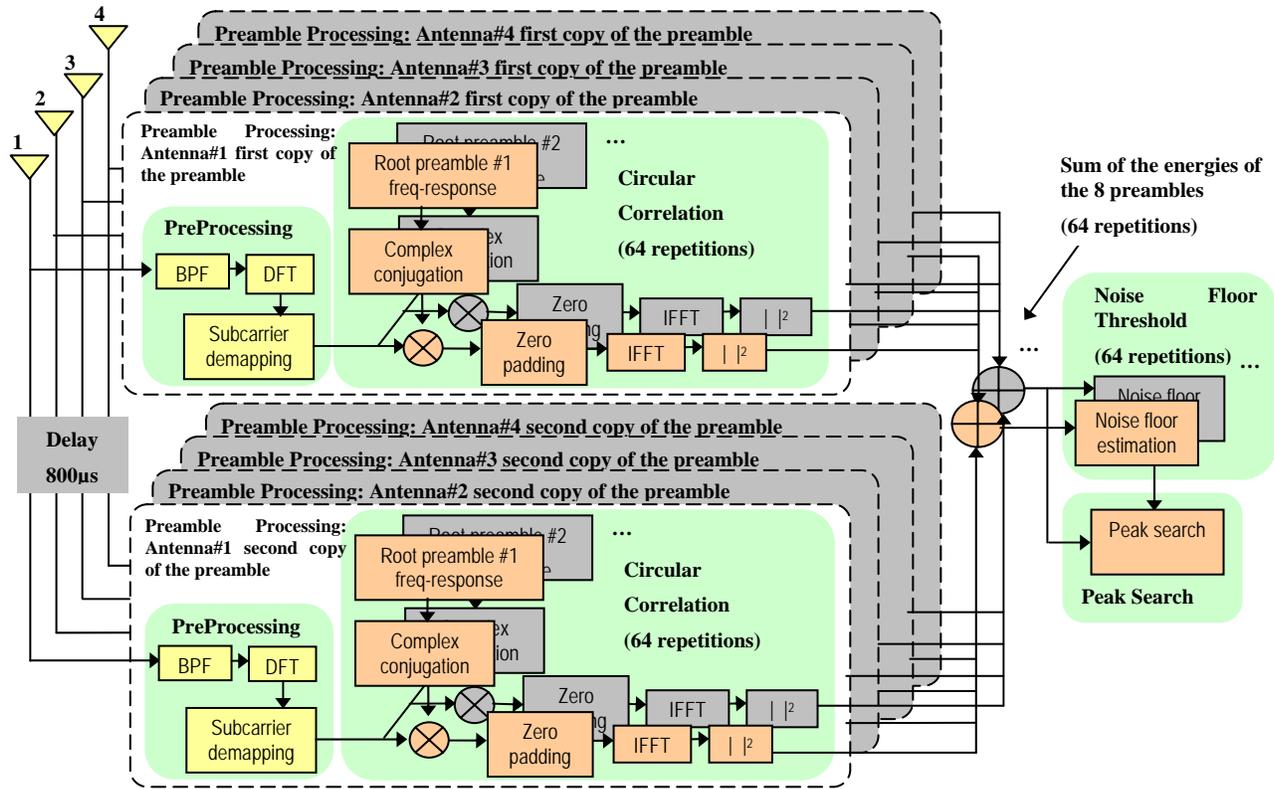

Fig. 1 Random Access Channel Preamble Detection (RACH-PD) Algorithm

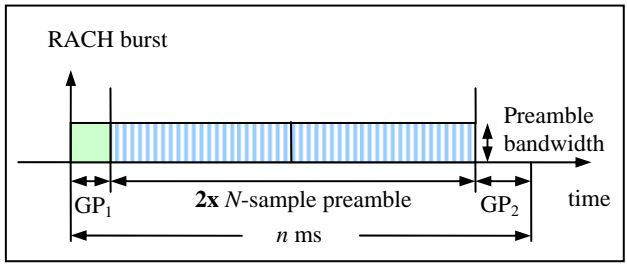

Fig. 2 A Random Access Slot Structure

The preamble is sent over a specified time-frequency resource, denoted as a *slot*, available with a certain cycle period and a fixed bandwidth. Within each slot, a guard period (GP) is reserved at each end to maintain time orthogonality between adjacent slots [1]. This preamble-based random access slot structure is shown in Figure 2.

The case study in this article assumes a RACH-PD for a cell size of 115 km. This is the largest cell size supported by LTE and also the case requiring the most processing power. According to [2], preamble format#3 is used with 21,012 complex samples as a cyclic prefix for GP1, followed by a preamble of 24,576 samples followed by the same 24,576 samples repeated. In this case the slot duration is 3 ms which gives a GP2 of 21,996 samples.

As per Figure 1, the algorithm for the RACH preamble detection can be summarized in the following steps [1]:

- After the cyclic prefix removal, the preprocessing (Preproc) function isolates the RACH bandwidth, by filtering with downsampling and then transforms the data into the frequency domain.
- Next, the circular correlation (CirCorr) function correlates data with several pre-stored preamble root sequences (or signatures) in order to discriminate between simultaneous messages from several users. It also applies an IFFT to return to the temporal domain and calculates the energy of each root sequence correlation.
- Then, the noisefloor threshold (NoiseFloorThr) function collects these energies and estimates the noise level for each root sequence.
- Finally, the peak search (PeakSearch) function detects all signatures sent by the users in the current time window. It additionally evaluates the transmission timing advance corresponding to the approximate user distance.

In general, depending on the cell size, three parameters of RACH may be varied: the number of receive antennas, the number of root sequences and the number of times the same preamble is repeated. The 115 km cell case displayed in Figure 1 implies 4 antennas, 64 root sequences, and 2 repetitions.

## III. THE ALGORITHM ARCHITECTURE MATCHING (AAM)

Currently, development tools for processors are primarily based on the C-language and an associated compilation tool. The major issue with a monolithic syntax is the inability to express parallelism. One solution is to use a Real-Time Operating System (RTOS) and to describe threads and their communication links (Mailboxes and pipes). Unfortunately, the application model used in an RTOS is too complex to handle multi-processor architectures when the number of threads increases [3]. For this reason, there is a need to explore methodologies better adapted at expressing the inherent parallelism within the application. Algorithm Architecture Matching (AAM [4]) is an example of one of these methodologies.

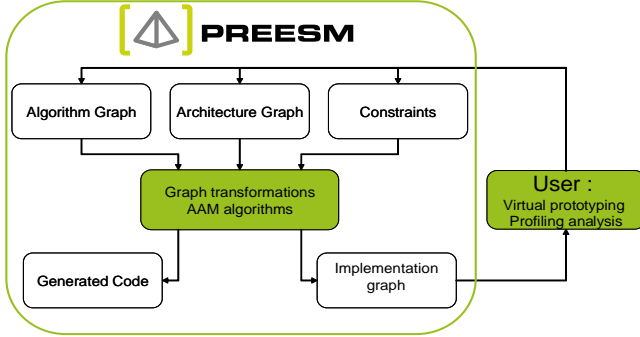

Fig. 3 PREESM Description

Algorithm Architecture Matching (AAM) maps an algorithm to a physical architecture given a set of constraints. The algorithm is described within PREESM using the algorithm graph (Figure 3). It relies on a description model which matches the application behavior. In the case of deterministic systems (including signal, image and communication applications), dataflow graphs have proven to be an efficient representation [5][6] for transformation-oriented systems and heterogeneous multi-component architectures. The algorithm graph (Figure 3) in PREESM is a Synchronous DataFlow graph (SDF) suitable for multi-processor architecture implementations [7]. Each vertex of the SDF represents an operation at coarse grain (equivalent of C function) and each edge represents a data dependency between the two operations at the end vertex. The vertices can be hierarchical, so allowing the description of the application at different resolutions. Thus, the SDF specifies the potential parallelism used in the matching step. The finest resolution vertex is called atomic operation; this type of operation may be described in a programming language such as C, VHDL, C++.

Within the PREESM tool, the architecture is described as the architecture graph (Figure 3) in which vertices represent operators and edges represent communication over a certain medium. An operator in this methodology is usually a processor connected to a local memory and has several communication resources. In this paper, operators are DSP cores and the media is an Enhanced Direct Memory Access (EDMA). The architecture graph specifies the available parallelism.

The matching consists of manually or automatically (AAM algorithms, Figure 3) exploring the implementation solutions with optimization heuristics. These heuristics aim to minimize the total execution time of the algorithm running on the multi-component architecture, by taking into account the execution time of operations and of data transfers between operations. The result of the matching allows automatic code generation [8] for multi-processor architectures handling synchronizations and data transfers between processors. Thus PREESM provides off-line static scheduling for multi-processor architectures. An implementation of AAM using the PREESM tool consists of:

- Performing a distribution (allocating parts of the algorithm to architecture components)
- Scheduling (determining the order for the operations distributed over a component) the algorithm on the architecture.
- Providing an implementation graph including simulation results of the distributed application functions.
- Generating C-code to verify the partitioning on target hardware and to provide a flexible implementation.

These functions enable PREESM to be used as an efficient virtual prototyping tool for our architecture exploration.

## IV. ARCHITECTURE EXPLORATION

### A. Algorithm Model

The goal of this exploration is to determine through simulation the architecture best suited to the 115km cell RACH-PD algorithm. The RACH-PD algorithm behavior is described as a SDF [3][9] in PREESM. An SDF description brings two major benefits to our implementation. The first is the proven possibility to schedule the algorithm statically. A static implementation enables static memory allocation, so removing the need for runtime memory administration. The second advantage is the high flexibility of communication parameter tuning, as achieved by modifying the SDF.

The RACH-PD algorithm model is shown in Figure 4. Initialization operations on the left-hand side are executed once as the system starts. Next, the three operations PreambleProcess, NoiseFloorThreshold and PeakSearch are executed sequentially in a loop while AntennaGen delivers samples to decode. The PreambleProcess operation is executed four times in each loop iteration; once per antenna. At the beginning of PreambleProcess, the atomic operation Preprocessing executes sequentially the bandpass filter, DFT and subcarrier demapping. It is repeated once for each of the two preamble repetitions. Then the circular correlation with 64 preamble root sequences is performed. Each circular correlation contains the correlation of the two preamble repetitions (SingleZCProc) with power accumulation similar to antenna power accumulation.

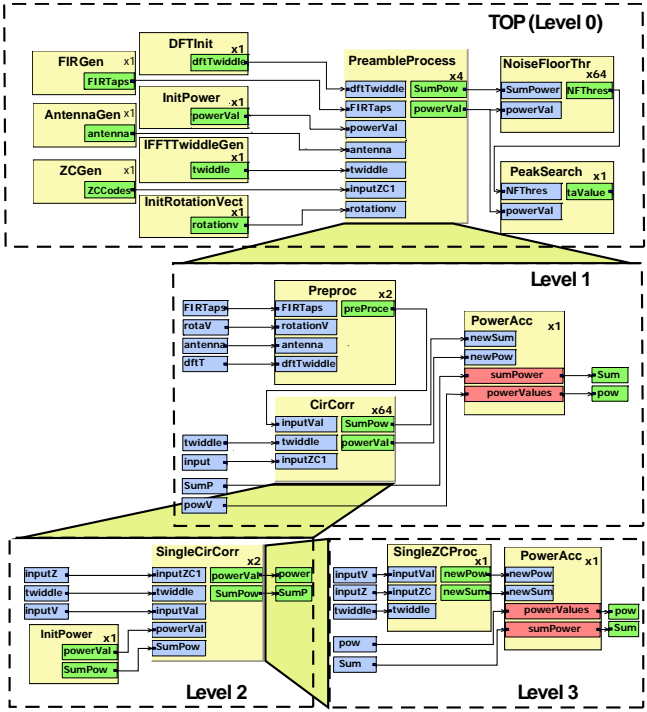

Fig. 4 Preamble Detection SDF Description

Using the same approach as in [10], valid scheduling derived from the representation in Figure 4 can be described by the compact expression:

$(8 \text{Pr}\,eproc)(4(64(InitPower(2((SingleZC\text{Pr}\,oc)(PowAcc))))PowAcc))$
$(64 NoiseFloorThreshold)PeakSearch$

We can separate the preamble detection algorithm in 4 steps:

- Preprocessing step: $8\text{Pr}\,eproc$
- Circular correlation step:
  $(4(64(InitPower(2((SingleZC\text{Pr}\,oc)(PowAcc))))PowAcc))$
- Noise floor threshold step: $(64 NoiseFloorThreshold)$
- Peak search step: $PeakSearch$

Each of these steps is mapped on the available cores and will appear in the exploration results detailed in Section IV-D. The given description generates 1,357 operations; this does not include the communication operations necessary in the case of multi-core architectures. Placing these operations by hand on the different cores would be greatly time-consuming. The architecture exploration PREESM tool offers an automatic scheduling, avoiding the problem of manual placement.

### B. Architecture Exploration

The four architectures explored are shown in Figure 5. The cores are all Texas Instrument TMS320C64x+ DSPs running at 1 GHz [11]. The connections are made via Direct Memory Access (DMA) links. The first architecture is a single-core DSP such as the TMS320TCI6482. The second architecture is dual-core, with each core similar to that of the TMS320TCI6482. The third is a tri-core and is equivalent to the new TMS320TCI6487 [12]. Finally, the fourth architecture is a theoretical architecture for exploration only, as it is a quad-core. The exploration goal is to determine the number of cores required to run the random RACH-PD algorithm in a 115 km cell and how to best distribute the operations on the given cores.

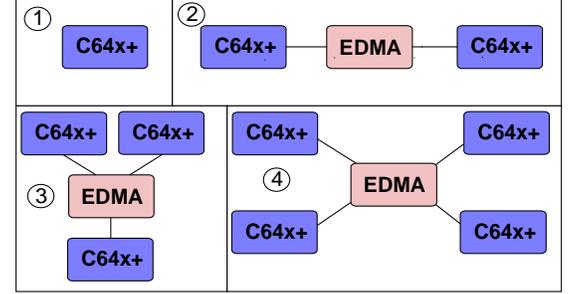

Fig. 5 Four architectures explored

### C. Architecture Model

To solve the implementation problem, each operation is assigned an experimental timing (in terms of CPU cycles). These timings are measured with implementations of the atomic functions on a single C64x+. The EDMA is modelled as a non-blocking medium transferring data at a constant rate. Assuming the EDMA has the same performance from the L2 internal memory to the L2 internal memory as the EDMA3 of the TMS320TCI6482, then the transfer of N bytes via EDMA should take approximately (see [13]):

$$transfer(N) = 135 + \frac{N}{3.375} cycles$$

The average size of the transmitted buffers in the 115 km preamble detection procedure is 4,800 bytes. Consequently, the average transfer speed used for simulation is 3.08 GBytes/s.

### D. Architecture Choice

The PREESM automatic scheduling process (i.e. the application of the AAM methodology to the RACH-PD algorithm) is applied for each architecture. The simulation results obtained are shown in Figure 6. Due to the 115 km cell constraints, preamble detection must be processed in less than 4 ms. Two kinds of experimental timings feed the simulation. The first set of timings is measured in loops, each calling a single function with L1 cache activated and appears as striped bars in Figure 6. It represents the application behaviour when data access is ideal. The second set of benchmarks is measured with L1 cache deactivated and leads to the higher cycles displayed in light grey. It represents the worst case of internal data accesses. For more details about C64x+ cache, see [11]. The RACH application is well suited for a parallel architecture, as the addition of one core reduces the latency dramatically. With L1 cache activated, two cores can process the algorithm within a time frame close to the real-time deadline. Simulation on the dual core with deactivated cache produces significantly higher cycles and misses the real-time deadline, so disqualifying the 2-core solution.

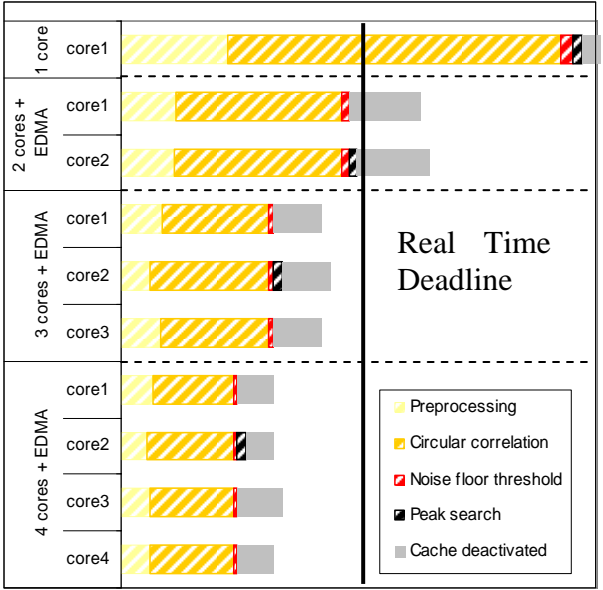

Fig. 6 Timings of the RACH-PD algorithm schedule on target architectures

The 3-core solution is clearly the best one: its CPU loads (68% with realistic cache misses and 88% without cache) are satisfactory and do not justify the use of a fourth core, as can be seen in Figure 6.

## V. IMPLEMENTATION ON THE CHOSEN ARCHITECTURE

With the architecture chosen, we can now start the static implementation process. Our goal is to automatically generate a highly optimized and flexible code with the necessary transfers and synchronization.

### A. Description of the chosen Architecture

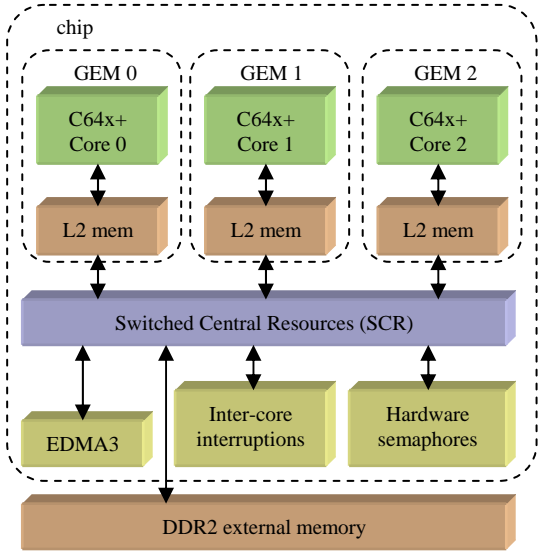

Fig. 7 Architecture of theTMS320TCI6487

The TMS320TCI6487 [12] is a three-core DSP specifically created for communication signal processing. Two modes are available for memory sharing: in symmetric mode, each CPU has 1MByte of L2 memory while in asymmetric mode, core 0 has 1.5Mbyte, core 1 has 1MByte and core 0.5MByte. Each CPU can access the L2 memory of the two other cores via the EDMA. Each CPU has also access to an external DDR2 memory. The EDMA can transfer a value from one core on-chip L2 memory to another core L2 memory in parallel with CPU calculation. This capability brings a higher flexibility than an architecture with cores interconnected via communication media.

Shared accesses between cores can be synchronized with hardware semaphores and inter-core interruptions. 32 semaphores may interrupt any core when a resource is accessed or released. Inter-core interruptions may launched from any core by writing in specific registers. Interruptions can carry a 7-bit value to distinguish one from another. Local to each CPU, the RTOS, DSP/BIOS, provides threads and local synchronization between threads with software semaphores. These features will be exploited to implement the RACH-PD algorithm and automatically generate function calls and synchronization. The use of software semaphores is consistent with a high performance implementation as passive wait is generated. Waiting for a DSP/BIOS semaphore puts the CPU in idle state.

### B. Using the EDMA as a message passing system

In order to prepare for code generation, we need to develop a communication library which provides synchronization. The communicator interface should be simple and may be called by generated code. The target architecture offers two communication possibilities: queues which are a message passing system built in DSP/BIOS operating system or the EDMA. As the queues are expected to be slower, the choice was made to use the EDMA.

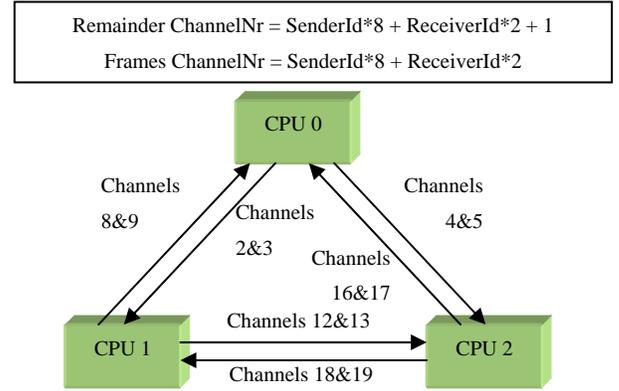

Fig. 8 EDMA channels used for communication between cores

The EDMA module offers 64 channels. Messages are split into N small frames and a remainder. The frames and the remainder are sent on two different chained channels. In order to avoid conflicts, 12 channels are used as shown in Figure 8. Since the channel number contains the sender and receiver identifiers, the receiver always knows, even in interruption routine, the sender of each communication received.

## C. Designing the Communication Process

Using the same method as in [6], the PREESM tool generates two threads per core: one for processing function calls and one for sending communication orders and waiting for transfer completion. As shown in Figure 9, when two successive functions are distributed on different CPUs, two semaphores Sem1 and Sem2 are generated on each core to synchronize the processing and communication threads. While communication threads are waiting for the completion of a transfer, processing threads can process data that does not impact this transfer. These local semaphores are implemented with DSP/BIOS operating system.

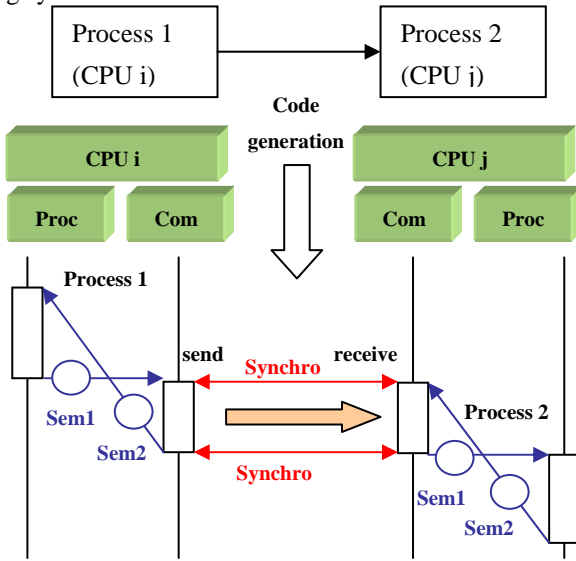

Fig. 9 Threads and synchronization within the cores

The next design problem to solve is the communication between cores. At the beginning of the communication process, the sender alone knows the source buffer address and the receiver alone knows the destination buffer address. There are two solutions (Figure 10) to complete a transfer in this situation: use an intermediate address or transfer the destination address.

When an intermediate address is used, the sender does not need to be aware of the destination address but each transfer must occur twice: from local memory to intermediate and from intermediate to destination. The dimension of the intermediate buffer is also a problem. In Figure 10, only the communication threads of the CPUs are represented.

The second solution is called memory pull because the receiver requests the data by sending its address. This solution imposes a bidirectional communication but is lighter than the preceding solution, as the address transfer of 4 bytes may be achieved through scratch buffers and synchronization through hardware semaphores or inter-core interruptions. During these transfers, we use only 12 of the 256 parameter set registers of the EDMA. The unused registers can be utilized as scratch buffers to transmit the transfer destination addresses.

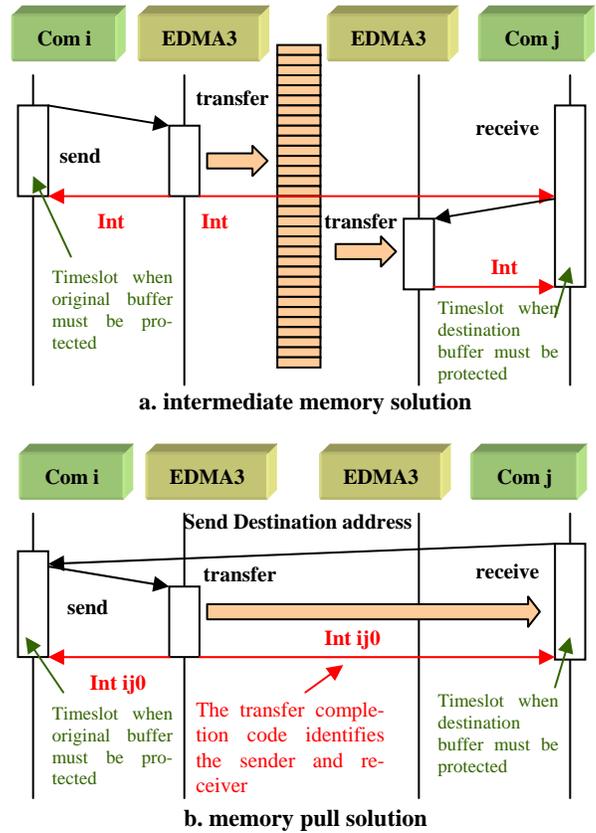

Fig. 10 Two solutions for the inter-core communication

The memory pull solution is chosen for two reasons. Firstly, the intermediate memory solution imposes the static allocation of an additional buffer of size at least as large as the largest one in the algorithm. In the RACH-PD case, this buffer should be of a minimum size of 100 kBytes with strong memory optimization. Secondly, the double transfer results in a division by 2 of the communication speed. Even with rates of several GBytes/s, this data rate reduction is not negligible. Simulation shows a communication cost of almost 5% for the RACH-PD algorithm.

As stated previously, two different modules may be used to generate inter-core synchronization: hardware semaphores or inter-core interruptions. In order to simplify the synchronization, a small library is created, with an interface close to the one of the local semaphores:
- HardSEMPend(int id): Waits for a semaphore with the given semaphore identifier
- HardSEMPost(int id): Launches a semaphore with the given semaphore identifier.

The semaphore identifier is chosen to be the Sender CPU number. Hardware semaphores are designed to protect critical sections from multiple accesses. They are typically used in resource access requests; access is granted when a semaphore is acquired. Programming the synchronization library with such a system leads to a clumsy implementation where the acquired resource is purely virtual. It is for this reason that the choice was made to base the library on inter-core interruptions. The principle of a inter-core synchronization library based on interruptions is quite simple. Any core can send an inter-core interruption to another core by providing the right identifier. An interruption is launched by the sender HardSEMPost function and is caught in an interrupt service routine of the receiver. The receiver then releases a local semaphore that was pending in the HardSEMPend function.

When the sender receives an interruption indicating the availability of the address, it launches the EDMA copy and waits. At the end of the transfer, the EDMA launches an interruption giving a transfer completion code equal to the EDMA channel number. The sender and receiver identifiers are deduced and communication threads of both CPUs are released.

When the cores are started, each core waits for the two other cores to run before sending any interruption. This system ensures that no inter-core interruption is ignored.

*D. RACH-PD memory consideration*

Once the communication model is complete, our attention turns to memory. The data buffers are statically allocated, some in the fast L2 memory of their CPU and others in the huge DDR2 external memory. We thus need to decide which buffers should be allocated in L2 and which configuration (symmetric or asymmetric) should be chosen. For the buffers in DDR2, sections of L2 memory must be used as cache for DDR2 so that performance does not decrease dramatically (see [14] for more details). Thus, we activate L2 cache with its maximal size of 256kBytes to improve DDR2 access time.

When the multi-core program is run with data in external memory, the EDMA reads and writes DDR2 data cached in L2. Cache coherency must then be taken into account. Indeed, the EDMA module runs the risk to read "dirty" data or to write in a cached value. Before sending data, a cache "write-back" is called to retrieve the data from cache. Before receiving data, a cache "invalidate" is called to mark the cache value as obsolete. With these precautions, cache coherency is maintained.

*E. Implementation tests*

When the communication programming is complete, we generate a 3-core code for C64x+ with the PREESM tool. We initially test a simple PREESM project by copying a buffer from one core L2 to another. The benchmarks of these copies are shown in Figure 11. The fixed overhead of an inter-core copy is approximately 2,700 cycles. This overhead is due to the synchronization process, the interruption routines and the EDMA configuration. When transferring big buffers, the EDMA data rate reaches 1.6 GBytes/s, half the speed of the TMS320TCI6482 EDMA employed in the simulations and benchmarked in [13].

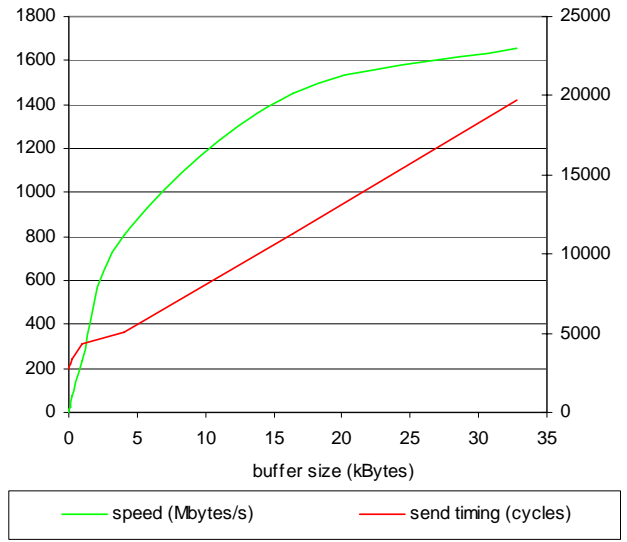

Fig. 11 Benchmarks of TCI6487 synchronized EDMA3 buffer copy

The RACH-PD algorithm is then tested and optimized on the TMS320TCI6487 platform in several steps:

*1) Single-core test*: A first version is tested on one core with code in L2 and data in DDR2. This implementation allows us to debug, dimension the stack and ensure that the code is working. The time of one preamble detection with this configuration is 240 ms.

*2) Cache activation*: L2 cache is then activated with the maximal size of 256 kBytes. The preamble detection time is then reduced to 69 ms.

*3) Three-core implementation*: The application is distributed over three cores. When we let the PREESM tool choose a strongly parallel implementation without constraints, it generates a very complex solution with 10,000 semaphores. We thus tune this solution, reducing inter-core communication and still allowing good pipelining (see Figure 12). The number of semaphores is then reduced to approximately 100. The complex solution which results from the non-constrained operation shows a limitation in the present PREESM mapping: a fixed cost for transfers needs be added to the tool to avoid the explosion of communication. The non-constrained PREESM automatic code generation allocates buffers of approximately 1.65 Mbytes for one core, 1.25 Mbytes for a second core and 200 kBytes for a third core, to which the heap and the code size must be added. This asymmetry justifies the use of asymmetric memory. With this configuration, one preamble detection takes 50 ms.

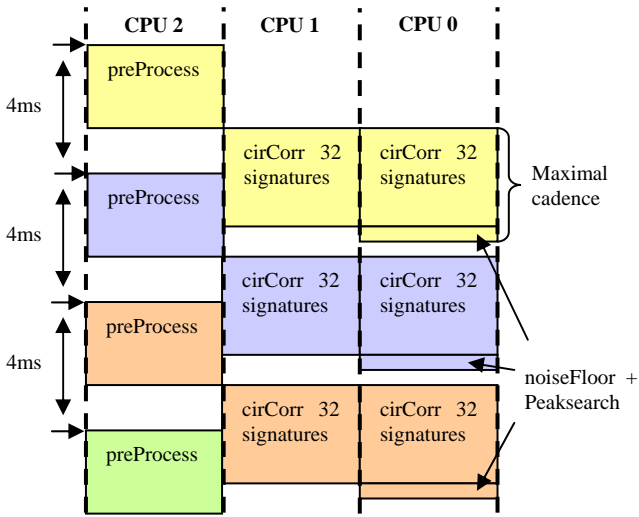

Fig. 12 Simple pipelining of the RACH-PD decoding

4) *Smart allocation and code*: If buffers are allocated in L2, pre-processing memory and code optimization brings the preamble detection processing time to approximately 10 ms. Internal allocation of some power buffers leads to a detection within 6.5 ms. Finally, after circular correlation buffer optimization, the detection time becomes 3.6 ms, under the constraint limit for response time.

The final implementation has one core loaded at 90%, one core at 75% and one at 70%. The simulation and code generation have led to a real implementation very close to that predicted with L1 cache worst case where the simulation loads were respectively: 88%, 83% and 83%. The added constraint of deactivating the cache in simulation inputs compensates for the external memory accesses that were not simulated and the EDMA rate slower than intended. These results show that prototyping the application enables a precise simulation of the multi-core solution before solving complex implantation problems.

## VI. FUTURE WORK

In the near future, a new communication model for the TMS320TCI6487 will be built based on message queues from DSP/BIOS operating system. This model will then be compared with that based on the EDMA. Its advantage will be the portability on devices using RapidIO (see [4]), as the operating system DSP/BIOS can use this communication system to pass messages.

Furthermore, other algorithms of LTE will be studied and implemented. It is expected that this will help to improve the architecture models of the PREESM tool. Specifically, internal/external allocation and advanced timings taking into account data caching may be automated, thus eliminating the manual step of memory allocation. Additionally, a more accurate communication model in the PREESM tool will remove the need for the manual reduction of semaphores.

## VII. CONCLUSIONS

The intent of this paper was to demonstrate a methodology using rapid prototyping and automatic code generation to develop an optimized multi-core implementation of a communication algorithm. After exploring 4 solutions, the best target architecture for the 115km cell RACH-PD algorithm was chosen, and an implementation is described and benchmarked. Memory allocation, function calls and EDMA calls are generated in C-code by the PREESM tool. Inter-core communication and memory partitioning are considered in the prototyping methodology. The result is an efficient and highly reconfigurable implementation, proving that the generation of static implementations from SDF descriptions is a viable solution for deterministic signal processing applications.

In the near future, an increasing number of CPUs will be available in complex System on Chips. Developing methodologies to efficiently partition code on these architectures is thus an increasingly important objective.